\documentclass[]{aa}
\usepackage{graphicx}

\begin{document}

\title{Discovery of a very cool object with extraordinarily strong
H$\alpha$ emission\thanks{These observations were collected at the VLT
of the ESO (Chile), the Keck I telescope (USA), and the 3.5-m
telescope at Calar Alto Observatory (Spain).}  }

   \author{David Barrado y Navascu\'es
          \and
          \inst{1}
          Mar\'{\i}a Rosa Zapatero Osorio
          \inst{1}
          \and
          Eduardo L$.$ Mart\'{\i}n
          \inst{2}
          \and
          V\'\i ctor J.\,S. B\'ejar
          \inst{3}
          \and
          Rafael Rebolo
          \inst{3,4}
          \and
          Reinhard Mundt
          \inst{5}
          }

\offprints{D. Barrado y Navascu\'es}

   \institute{Laboratorio de Astrof\'{\i}sica Espacial y F\'{\i}sica 
        Fundamental,
           INTA,  P.O. Box 50727, E--28080 Madrid, Spain\\
              \email{barrado@laeff.esa.es}
         \and
             Institute of Astronomy. University of Hawaii at Manoa. 
             2680 Woodlawn Drive, Honolulu, HI 96822, USA.
         \and
             Instituto de Astrof\'\i{}sica de Canarias, E--38205 La 
             Laguna, Tenerife, Spain
         \and
             Consejo Superior de Investigaciones Cient\'{\i}ficas, CSIC, 
             Spain
         \and
             Max-Planck-Institut f\"ur Astronomie,
             K\"onigstuhl 17, D--69117 Heidelberg, Germany 
             }

   \date{Received; accepted }

   \abstract{ We report on the finding of the  strongest
H$\alpha$ emission -- pseudoequivalent width of 705~\AA~-- known so
far in a young, late type dwarf. This object, named as S\,Ori~71, is a
substellar candidate member of the 1--8\,Myr star cluster
$\sigma$\,Orionis. Due to its overluminous location in
color-magnitude diagrams, S\,Ori~71 might be younger than other
cluster members, or a binary of similar components.  Its mass is in
the range 0.021--0.012\,$M_\odot$, depending on evolutionary models and
possible binarity. The broad H$\alpha$ line of S\,Ori~71 appears
asymmetric, indicative of high velocity mass motions in the H$\alpha$
forming region. The origin of this emission is unclear at the present
time. We discuss three possible scenarios: accretion from a disk, mass
exchange between the components of a binary system, and emission from
a chromosphere.  
 }

  \titlerunning{H$\alpha$ in a very cool dwarf}

  \authorrunning{Barrado y Navascu\'es et al.}

   \maketitle

%
%


\section{Introduction}

Both mass accretion and chromospheric activity have H$\alpha$ emission
as a signature.  Chromospheric activity appears as a consequence of
the low density, the inverted temperature profile and the strong
magnetic field present in K- and M-type dwarfs. Other Balmer lines, as
well as Ca{\sc ii} H\&K and the Ca{\sc ii} infrared triplet can appear
in emission too.  The activity detected in some brown dwarfs (objects
incapable of burning hydrogen stably, with masses below 0.075 $M$$_\odot$,
Chabrier at al$.$ \cite{chabrier00} and references therein) have,
probably, this origin (Mart\'{\i}n et al$.$ \cite{martin99a}),
although so far there is no theoretical model capable of explaining
this phenomenology (e.g., Mohanty et al$.$ \cite{mohanty02}). In
addition, the H$\alpha$ emission line of these cool objects often has
an intrinsic variability, and strong flares are commonly detected in M
dwarfs.  On the other hand, accretion can appear in interacting
binaries (during the transfer process of material) or in accreting
objects with circumstellar disks, such as the young TTauri
stars. These pre-main sequence stars are normally classified as
classical TTauri (CTT) stars or weak-line TTauri (WTT) stars.  The
first group is characterized by strong H$\alpha$ emission (larger than
10 or 20 \AA, Appenzeller \& Mundt \cite{appenzeller89}; Mart\'{\i}n
\cite{martin97}), asymmetric and broad H$\alpha$ profiles (sometimes
with double peaks), presence of forbidden emission lines (arising from
shocks produced by jets and outflows), blue/UV and infrared excesses,
and a strong Li\,{\sc i}\,6708\,\AA~line in absorption (indicative of
youth). On the contrary, WTT stars lack most of these properties due
to the absence of an active disk, and show smaller H$\alpha$ emissions
while keeping strong lithium lines. In addition, WTT stars display a
lower degree of variability (e.g., Herbst et al$.$ \cite{herbst02}).
 
The $\sigma$\,Orionis cluster (Walter et al$.$ \cite{walter97}) is a
young (1--8\,Myr, Zapatero Osorio et al$.$ \cite{osorio02a}) stellar
association with low reddening ($E(B-V)$=0.05, Lee \cite{lee68}), and
located at a Hipparcos distance of 352$^{+166}_{-85}$\,pc. Many
substellar members of this cluster show H$\alpha$ emission in their
optical spectra (Barrado y Navascu\'es et al$.$ \cite{barrado02}, and
references therein).  This paper presents the discovery of a brown
dwarf with likely membership in $\sigma$\,Orionis, which has the
strongest H$\alpha$ emission line ever detected in a TTauri late-type
star or active very cool object.  We discuss the origin of this
extraordinary emission.

\setcounter{table}{0}
\begin{table*}
\caption[]{ Photometric and spectroscopic data of S\,Ori~71
(S\,Ori\,J053900.2--023706).  Photometric errors: $\pm$\,0.10\,mag. }
\begin{tabular}{ccccccccc}
\hline
RA (J2000) DEC                            & $I_c$ &$I_c-J$& 
SpT.            & pW (H$\alpha$) & $T_{\rm eff}$ & 
log\,$L/L_{\odot}$ & Mass (single) & Mass (binary) \\
($^{h\,m\,s}$)\hspace{10mm}($^\circ$ ' '') &      &       &
                & (\AA)          & (K)           &
                   & ($M_{\odot}$) & ($M_{\odot}$) \\
\hline
05 39 00.2 ~ --02 37 06                   & 20.02 & 2.88  & 
L0\,$\pm$\,0.5  & 705\,$\pm$\,75 & 2200--2500    &
--2.66\,$\pm$\,0.15& 0.014--0.021  & 0.012--0.017  \\ 
\hline	
\end{tabular}

\end{table*}

\section{Observations and analysis}

The object studied in this paper, named S\,Ori~71, was discovered in
an optical $IZ$ survey using LRIS at the 10-m KeckI telescope, USA
(Zapatero Osorio et al$.$ \cite{osorio02b}).  Near-infrared data
($J$-band) were collected with the Omega Prime detector attached at
the 3.5-m telescope of the Calar Alto Observatory, Spain.  The
combination of optical and infrared datasets produced several brown
dwarf and isolated planetary-mass candidates (unable to fuse
deuterium) of the $\sigma$\,Orionis cluster. Many of them have been
followed-up spectroscopically in order to assess their nature and
study their properties (Barrado y Navascu\'es et al$.$
\cite{barrado01}; Mart\'\i n et al$.$ \cite{martin01b}).  The 
spectrum presented in this paper was collected with the 8-m VLT/UT1
telescope at the Paranal Observatory of the European Southern
Observatory, Chile, using FORS1 (2451904.056~JD).  It was processed
and analyzed as in Barrado y Navascu\'es et al$.$
(\cite{barrado01}). The VLT spectrum of S\,Ori71 (6200--9300\,\AA) is
shown in Fig.~1. The spectral type of our object was derived by
comparing its pseudo-continuum to that of various M- and L-type dwarf
spectra, and by comparing several spectral ratios in different bands
following Kirkpatrick et al$.$ (\cite{kirk99}) and Mart\'{\i}n et
al$.$ (\cite{martin99b}).  The final classification, L0, has a
uncertainty of half a sub-class.  Coordinates, photometric and
spectral measurements, and other parameters of S\,Ori\,71 are listed
in Table~1. A finding chart can be provided upon request.  Both
photometric and spectroscopic data support S\,Ori\,71's being a likely
member of the $\sigma$\,Orionis cluster.

   \begin{figure}
   \centering
   \includegraphics[width=9cm]{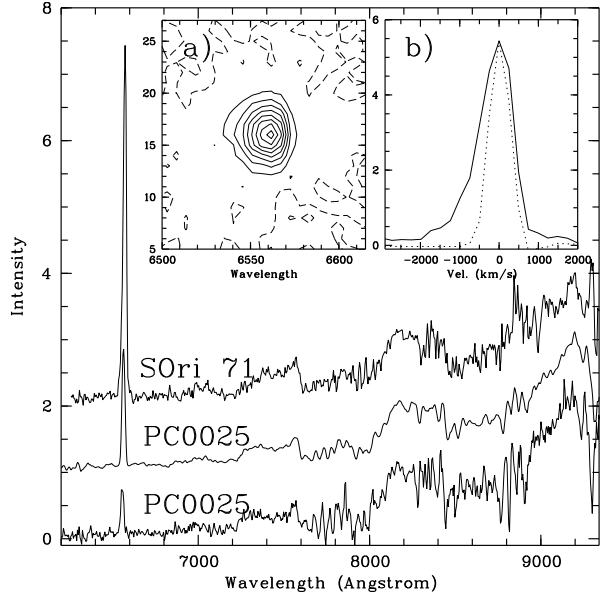}
\vspace{-4.8cm}
\caption{Spectra of S\,Ori\,71 (this paper) and PC\,0025$+$0047.
  Two enlargements of the
H$\alpha$ line of S\,Ori\,71 are also shown.  {\bf a)} Contour diagram
for the 2-D spectrum.  Note the asymmetry.  Dashed lines correspond to
zero level, whereas the peak of the H$\alpha$ line is at 431 counts.
Each contour corresponds to a step of 50 counts.  {\bf b)} 1-D
spectrum -- solid line -- and a comparison with the instrumental
profile -- dotted line.  }
\end{figure}

   \begin{figure}
   \centering
   \includegraphics[width=9cm]{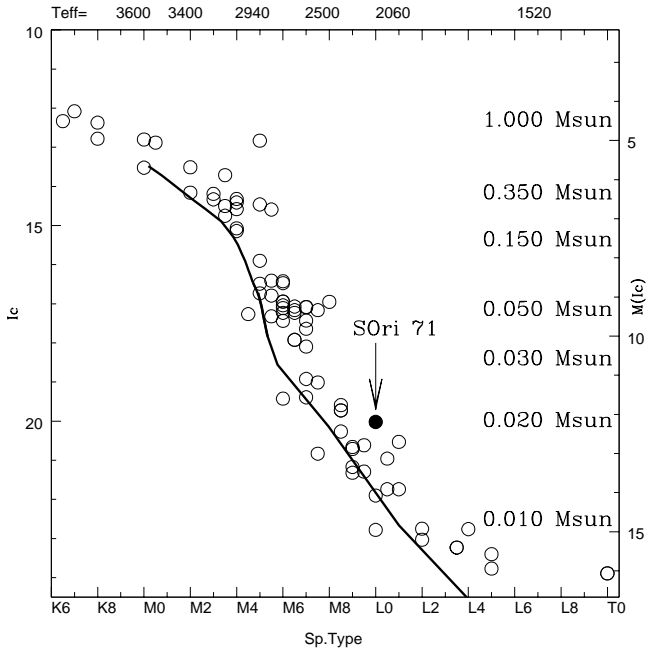}
\caption{ Spectral type against $I_c$ magnitude of members of the
$\sigma$\,Orionis cluster.  Data from Barrado y Navascu\'es et al$.$
(\cite{barrado01}, \cite{barrado02}), Mart\'{\i}n et al$.$
(\cite{martin01b}), B\'ejar et al$.$ (\cite{bejar01}) and Zapatero
Osorio et al$.$ (\cite{osorio02a}).  The line represents a
3\,Myr-isochrone from Baraffe et al$.$ (\cite{baraffe98}). Absolute
magnitudes and masses are indicated at the right hand side of the
diagram, and effective temperatures (in K) are shown at the top.}
\end{figure}


Figure~2 illustrates the known spectral sequence of the
$\sigma$\,Orionis cluster. S\,Ori\,71 lies above this sequence, which
might indicate that, either it is younger than other cluster members,
or that it contains two components of rather equal mass. A similar
case is that of S\,Ori\,47 (Zapatero Osorio et al$.$ \cite{osorio99};
Barrado y Navascu\'es et al$.$ \cite{barrado01}), a slightly cooler
(L1), substellar object of the cluster.  S\,Ori\,71 also appears
overluminous in color-magnitude diagrams. Its luminosity (Table~1) is
obtained using bolometric corrections of Leggett et al$.$
(\cite{leggett02}). Adopting a cluster age of 3\,Myr and the
evolutionary models of Baraffe et al$.$ (\cite{baraffe98}), we
estimate the mass of S\,Ori\,71 to be between 0.014\,$M_\odot$ and
0.021\,$M_\odot$ if it were a single object. In case of binarity, each
component would be 0.012--0.017\,$M_\odot$. From the models, we also
infer that the surface temperature of the object is in the range
2200--2500\,K, in agreement with its spectral type.  Table~1
summarizes these results.  Other sets of theoretical models (Burrows
et al$.$ \cite{burrows97}; D'Antona \& Mazzitelli \cite{dantona97};
Chabrier et al$.$ \cite{chabrier00}) produce similar mass and temperature values.

\section {H$\alpha$ emission and its origin}

S\,Ori\,71 displays a broad, incredibly strong H$\alpha$ emission as
compared to other similar type objects. Two blow-ups of the object's
spectrum around H$\alpha$ are depicted in Fig.~1: panel $a$
illustrates the asymmetry of the line in a 2-D plot, while the
comparison to the instrumental profile (sky line) is provided in panel
$b$.  This broad H$\alpha$ emission is probably a result of the
electron scattering in the H$\alpha$ formation region as observed in
some TTauri stars (Stahl \& Wolf \cite{stahl80}).  Figure~1 also
includes two spectra of the field dwarf PC\,0025$+$0047 taken from
Mart\'{\i}n et al$.$ (\cite{martin99a}). With M9.5 spectral type, this
field object shows a persistent and variable H$\alpha$ emission. The line
pseudo-equivalent widths (pEWs) of the two spectra of the figure are
$\sim$200\,\AA~and 400\,\AA, although Mart\'{\i}n et al$.$
(\cite{martin99a}) have reported H$\alpha$ variability in the range
100--400\,\AA~in a 4\,yr timespan.  Other few field objects with
similar spectral types also have strong H$\alpha$ emissions up to
pEW\,$\sim$\,400\,\AA~(Liebert et al$.$ \cite{liebert99}; Burgasser et
al$.$ \cite{burgasser02}).  However, the emission of S\,Ori~71
(pEW\,=\,705\,$\pm$\,75\,\AA) stands out due to its huge intensity.
Table~2 summarizes the properties of several objects with similar
spectral classes both in the field and in the $\sigma$\,Orionis
cluster. To our knowledge, in addition to these objects, the largest
H$\alpha$ pEWs correspond to more massive and warmer sources, like
LkH$\alpha$~101, an unusual F-type CTT star (Cohen \& Kuhi
\cite{cohen79}; Herbig \& Bell \cite{herbig88}), and XZ~Tau, V573~Ori,
HO~Lup, Sz~123, Sz~69, Sz~102, WZ~Cha and PT~Mon, all of them with K
and early-M classes and H$\alpha$ pEWs\,=\,220--377\,\AA.

We have also compared the H$\alpha$ pEWs of $\sigma$\,Orionis stellar
and substellar cluster members (Bej\'ar et al$.$ \cite{bejar99};
Barrado y Navascu\'es et al$.$ \cite{barrado01}; Zapatero et al$.$
\cite{osorio02a}, \cite{osorio02c}) to those values of CTT stars of
Orion with less than 1\,Myr (Herbig \& Bell \cite{herbig88}), WTT
stars (Alcal\'a et al$.$ \cite{alcala96}, \cite{alcala00}), and very
low mass stars and brown dwarfs of the $\alpha$\,Persei cluster
(70--95\,Myr, Prosser et al$.$ \cite{prosser94}; Stauffer et al$.$
\cite{stauffer99}). Fine details are given in Barrado y Navascu\'es et
al$.$ (\cite{barrado02}). While H$\alpha$ emission in CTT stars is
very likely originated in accretion disks surrounding the central
object, older stars are believed to show less intense H$\alpha$
emission as a consequence of chromospheric activity and
rotation. Resulting from the comparison, we note that besides the very
young age of the $\sigma$\,Orionis cluster, Orion CTT stars have
stronger H$\alpha$ lines, and that the emission level of the
$\sigma$\,Orionis low mass members is slightly larger or similar to
that of the $\alpha$\,Persei cluster (late-K to mid-M spectral
types). This late cluster is older (Stauffer et al$.$
\cite{stauffer99}; Basri \& Mart\'{\i}n \cite{basri99}), but its low
mass members have high rotation rates (Randich et al$.$
\cite{randich96}). On the other hand, H$\alpha$ pEWs of the
$\sigma$\,Orionis cluster members, regardless of its origin, increases
toward cooler objects, reaching about 100\,\AA~in the planetary-mass
domain (Barrado y Navascu\'es et al$.$ \cite{barrado01}). Line
variability is also present among the smallest objects: Zapatero
Osorio et al$.$ (\cite{osorio02c}) measured a variable emission in
S\,Ori\,55 (Table~2), with pEWs ranging between 180\,\AA~and 410\,\AA,
whereas few weeks before, Barrado y Navascu\'es et al$.$
(\cite{barrado01}) reported almost no activity
(pEW\,=\,5~\AA). Unfortunately, we have a single spectrum of
S\,Ori\,71, and cannot reach any conclusion on its variability.
However, we remark that S\,Ori\,71 presents a noteworthy H$\alpha$ pEW
among low mass stellar and substellar members of any young
cluster. Our literature search covered objects in Upper Scorpius, the
Scorpius-Centaurus complex, Taurus, and the IC\,348 cluster (Cohen \&
Kuhi \cite{cohen79}, Herbig \& Bell \cite{herbig88}; Alcal\'a et al$.$
\cite{alcala96}, \cite{alcala00}; Mart\'{\i}n \cite{martin98}; Herbig
\cite{herbig98}; Ardila et al$.$ \cite{ardila00}).

\setcounter{table}{1}
\begin{table}
\caption[]{Properties of S\,Ori\,71 and other similar type objects. }
\begin{tabular}{lllll}
\hline
Property              & S\,Ori\,71 & S\,Ori\,55 & PC\,0025     & LHS\,2065 \\  
\hline			       		          
Sp$.$ Type            & L0         & M9         & M9.5         & M9        \\ 
Age    (Myr)          & 2--8       & 2--8       & $\le$600     & ?         \\ 
Mass ($\times10^{-3}\,M_\odot$)
                      & 21--12     & 16--8      & $\le$70      & $\ge$65   \\ 
Lithium               & ?          & ?          & Y            & N         \\ 
pW(H$\alpha$) (\AA)   & 705        & 5--410     & 100--400     & 7.5--261  \\  
H$\alpha$ asymmetry   & Y          & N?         & N            & N         \\ 
H$\alpha$ variability & ?          & Y          & Y            & Y         \\ 
Flares                & ?          & Y          & N            & Y         \\ 
Forbidden lines       & N?         & N?         & N?           & N         \\ 
He{\sc i} 6678\,\AA   & N          & N          & N            & Y         \\
Optical veiling       & N?         & N?         & Y            & N         \\ 
IR excess             & ?          & ?          & N            & N         \\ 
\hline
\end{tabular}

PC\,0025 (Mart\'{\i}n et al$.$
\cite{martin99a}); LHS\,2065 (Mart\'{\i}n \& Ardila \cite{martin01a});
S\,Ori\,55, cluster member (Zapatero Osorio et al$.$
\cite{osorio02c}).
\end{table}

If we consider the ratio between the H$\alpha$ luminosity of the
object to its bolometric luminosity, as shown in Fig.~3, S\,Ori\,71
appears to be more active than any other known brown dwarf
(log\,$L(\rm H\alpha)$/$L_{\rm bol})$\,=\,--2.69\,$\pm$\,0.16). In
very young star forming regions, strong H$\alpha$ emission is
generally due to disk accretion, and it may be accompanied by emission
of forbidden lines, like [O\,{\sc i}] at 6300\,\&\,6364\,\AA, [N\,{\sc
ii}] at 6548\,\&\,6583\,\AA, and [S\,{\sc ii}] at 6717\,\&\,6731\,\AA,
which result from jets and outflows.
Other lines indicating activity, such as He\,{\sc i} at 6678\,\AA,
are not seen.
 We can impose upper limits of
1--2\,\AA~to the pEW of these lines from our spectrum.

 The fact that S\,Ori\,71 could be a binary formed by
similar mass objects, as previously discussed, adds a new possibility
to the origin of the H$\alpha$ emission. Some matter transfer might
take place between the components if they are close enough to each
other (as proposed by Burgasser et al$.$ \cite{burgasser00} for a T
spectral type brown dwarf that has a persistent line
emission). However, more data are needed to assess the reliability of
this suggestive scenario. In addition, we have considered the
possibility of S\,Ori\,71 being an interacting binary formed by one
compact object and a very small object. In this case, S\,Ori\,71 would
not belong to the $\sigma$\,Orionis cluster. In order to reproduce the
observed photometric data of our object, it would be composed of a
0.09\,$M_\odot$ main sequence star and a $\sim$0.5\,$M_\odot$ white
dwarf ($\tau \ge$\,2\,Gyr) in a very short orbit. However, our
optical spectrum does not show any evidence for a contribution from a
hot companion (the white dwarf).

   \begin{figure}
   \centering
   \includegraphics[width=9cm]{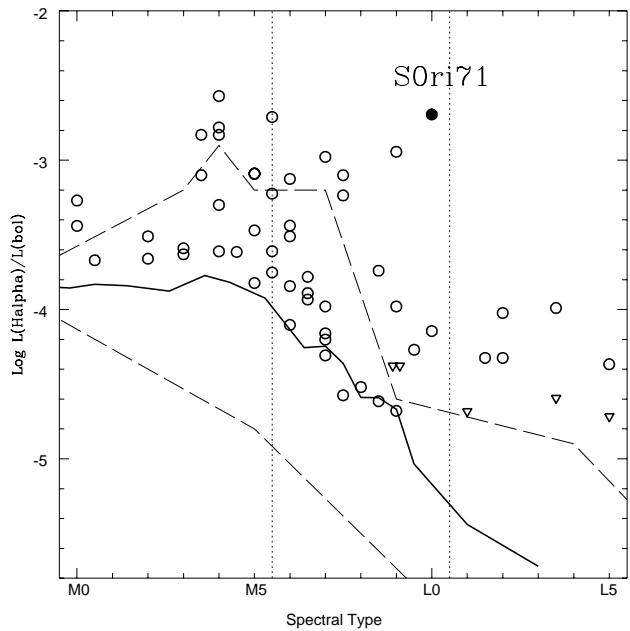}
\caption{ Ratio between H$\alpha$ and bolometric luminosities as a
function of spectral type for $\sigma$\,Orionis members (open circles;
open triangles denote upper limits). The solid-thick line stands for
the average for field objects, whereas the area within the dashed
lines indicates the region occupied by individual measurements of M-
and L-type field dwarfs (Burgasser et al$.$ \cite{burgasser02}).
 Note that most of the L
field dwarfs only have upper limits. The vertical dotted lines
separates the stellar (left), brown dwarf (centre) and planetary mass
(right) regimes in $\sigma$\,Orionis.}
\end{figure}

It might be possible to classify S\,Ori\,71 as a a CTT substellar
analog (i.e., accreting from a surrounding disk), based both on the
strength of H$\alpha$ and its asymmetric profile. It would be very
interesting to study the object's $K$-band and mid-infrared emission
and see if there is any flux excess, which would confirm the presence
of the disk. At the present time and with the available data, we
cannot rule out any possible scenario. A further explanation would be
the case of an interacting binary with mass exchange or a common
chromosphere/corona where magnetic loops would connect both objects
and would experience frequent reconnections, producing the very strong
H$\alpha$ emission. The geometry of these loops would be distorted as
both components orbit around each other (specially if there is not
synchronization between the orbital and the rotational periods),
producing a magnification of the magnetic fields by compressing the
magnetic lines, and resulting in abrupt releases of energy during
flare-like events.  Additional infrared, X-ray and radio data would be
very valuable to provide new insights on the origin of the H$\alpha$
emission of objects like S\,Ori\,71.

\begin{acknowledgements}
We thank the ESO staff at Paranal.  Financial support was provided by
 the Spanish {\it ``Programa Ram\'on y Cajal''} and AYA2001-1124-CO2
 programs.
\end{acknowledgements}

\end{document}